\documentclass[preprint,showpacs,showkeys,aps,prb]{revtex4}
\usepackage[english]{babel}
\usepackage[dvips]{graphicx}

\begin{document}

\title{
Ergodicity of the Martyna-Klein-Tuckerman Thermostat \\
and the 2014 Snook Prize
} 

\author{
{\bf William Graham Hoover} and {\bf Carol Griswold Hoover}, \\
Ruby Valley Research Institute \\
Highway Contract 60 Box 601 \\
 Ruby Valley Nevada USA 89833 ; \\
}

\date{\today}



\vspace{0.1cm}

\begin{abstract}
Nos\'e and Hoover's 1984 work showed that although Nos\'e and Nos\'e-Hoover
dynamics were both consistent with Gibbs' canonical distribution neither
dynamics, when applied to the harmonic oscillator, provided Gibbs' Gaussian
distribution.  Further investigations indicated that two independent thermostat
variables are necessary, and often sufficient, to generate Gibbs' canonical
distribution for an oscillator.  Three successful time-reversible and
deterministic sets of two-thermostat motion equations were developed in the
1990s. We analyze one of them here.  It was developed by Martyna, Klein, and
Tuckerman in 1992. Its ergodicity was called into question by Patra and
Bhattacharya in 2014.  This question became the subject of the 2014 Snook
Prize.  Here we summarize the previous work on this problem and elucidate
new details of the chaotic dynamics in the neighborhood of the two fixed points.
We apply six separate tests for ergodicity and conclude that the MKT equations
are fully compatible with all of them, in consonance with our recent work with
Clint Sprott and Puneet Patra.
\end{abstract}

\maketitle

\section{Deterministic Time-Reversible Thermostats and Ergodicity}

In 1984 Shuichi Nos\'e discovered a canonical dynamics\cite{b1,b2} consistent
with Willard Gibbs' canonical phase-space distribution.  Hoover\cite{b3} used a
generalization of Liouville's flow equation to develop a ``Nos\'e-Hoover dynamics'',
a simpler variation on Nos\'e's work.  He pointed out that neither approach gave
Gibbs' complete canonical distribution for the simple harmonic oscillator problem :
$$
{\cal H}(q,p) = [ \ (q^2/2) + (p^2/2) \ ] \longrightarrow
f(q,p) \propto e^{-q^2/2}e^{-p^2/2} \ [ \ {\rm Gibbs} \ ] \ . 
$$
In the 1990s three generalizations of this work were developed\cite{b4,b5,b6} to
remedy the stiffness and the lack of ergodicity that resulted when Nos\'e's ideas
were applied to the harmonic oscillator.  All three of them include two ``thermostat''
variables, $\{ \ \zeta,\xi \ \}$ , which control the motion of the oscillator
coordinate and momentum $\{ \ q,p \ \}$ , steering it toward the canonical
distribution. The generalized motion equations all satisfy an analog of the
hydrodynamic continuity equation, $(\partial \rho/\partial t) = -\nabla \cdot (\rho u)$ .
The stationary ( steady-state ) version of this phase-space flow equation is :
$$
(\partial f/\partial t) = 0 = -(\partial f\dot q/\partial q) -(\partial f\dot p/\partial p)
-(\partial f\dot \zeta /\partial \zeta)-(\partial f\dot \xi/\partial \xi) \ .
$$
For all three flow models the probability density $f(q,p,\zeta,\xi)$ is stationary when
it includes {\it Gaussian} distributions for the two new thermostat variables, $\zeta$
and $\xi$ .

As a result there are at present three sets of four ordinary differential
motion equations all of which provide the full canonical distribution for an
oscillator along with Gaussian distributions for the additional thermostat
variables $\{ \ \zeta,\xi \ \}$ :
$$
\{ \ \dot q = p \ ; \ \dot p = -q - \zeta p -\xi p^3 \ ; \ \dot \zeta = p^2 - 1 \ ; \
\dot \xi = p^4 - 3p^2 \ \} \ [ \ {\rm HH} \ ]\cite{b4} \ ;
$$
$$
\{ \ \dot q = p \ ; \ \dot p = -q - \zeta ^3p -\xi p^3 \ ; \ \dot \zeta = p^2 - 1 \ ; \ 
\dot \xi = p^4 - 3p^2 \ \} \ [ \ {\rm JB} \ ]\cite{b5} \ ;
$$
$$
\{ \ \dot q = p \ ; \ \dot p = -q - \zeta p \ ; \ \dot \zeta = p^2 - 1 - \xi \zeta \ ; \ 
\dot \xi = \zeta^2 - 1 \ \} \ [ \ {\rm MKT} \ ]\cite{b6} \ .
$$
Each of them displays a ``mirror'' or ``inversion'' or ``rotational'' symmetry in the
$(q,p)$ plane: any solution $\{ \ +q(t),+p(t),\zeta(p),\xi(t) \ \}$ has a mirror image
when the oscillator is viewed in a mirror perpendicular to the $q$ axis.  The solution
viewed in the mirror replaces both $+q$  and $+p$ by their mirror images, $-q$ and $-p$ .
In the mirror solution, $\{ \ -q(t),-p(t),\zeta(p),\xi(t) \ \}$ , the time-dependent
thermostat variables $\zeta$ and $\xi$ are unchanged.

There are also generalizations of each of these ideas based on controlling more, or
different, moments of the canonical distribution function.\cite{b7,b8}  In addition,
a variety of different solutions result for thermostat relaxation times other than
unity, for coordinate-dependent temperature profiles, and for more complicated potentials.

In addition to the canonical oscillator probability $\propto e^{-q^2/2}e^{-p^2/2}$ the
thermostat variables $(\zeta,\xi)$ also have Gaussian distributions :
$$
f_{HH} = f_{MKT} \supset e^{-\zeta^2/2}e^{-\xi^2/2} \ ;
\ f_{JB}  \supset e^{-\zeta^2/2}e^{-\xi^4/4} \ .
$$

Patra and Bhattacharya\cite{b9} investigated the $(q,p)$ phase-space density in the
vicinity of an unstable fixed point $(q,p,\zeta,\xi)=(0,0,-1,+1)$ of the MKT equations.
They displayed an apparent low-probability region there and suggested that the MKT
equations were not ergodic. Because any lack of ergodicity would contradict Martyna,
Klein, and Tuckerman's belief in the ergodicity of their own model, we established
the 2014 Snook Prize\cite{b10} as a reward for the most convincing work  demonstrating
either ergodicity or its lack.  In January 2015 we awarded the prize to the authors of
Reference 11.

Here we clarify the differing conclusions of References 9 and 11 by exploring six aspects
of the chaotic dynamics and stationary measure of the MKT equations. These include [A]
the moments of the measure, [B] the largest Lyapunov exponent, [C] the two fixed points
of the flow, [D] the attractor/repellor dynamics near the two fixed points, [E] the
measure in the neighborhood of these fixed points, and [F] the symmetry of the measure
in the neighborhoods of 81 lattice points arranged in a four-dimensional phase-space
lattice. Our description of the underlying analysis and numerical work makes up the
following Section II.  Our Conclusions follow in the Summary Section III.

\section{Investigating Ergodicity: the MKT Harmonic Oscillator}

Ergodicity, with any dynamical trajectory coming close to {\it all} phase-space states,
became an issue with the study of the one-thermostat Nos\'e-Hoover oscillator\cite{b3,b12} :
$$
\dot q = p  \ ; \ \dot p = -q - \zeta p \ ; \ \dot \zeta = p^2 - 1 \ .
$$
This model exhibits a variety of regular solutions. Most trajectories correspond to
two-dimensional tori in the three-dimensional $(q,p,\zeta)$ phase-space.  About five
percent of the Gaussian phase-space measure ,
$$
(2\pi)^{3/2}f(q,p,\zeta) = e^{-q^2/2}e^{-p^2/2}e^{-\zeta^2/2} ,
$$
makes up a chaotic sea perforated by the tori\cite{b11}.  

Surprisingly, adding a fourth variable to the phase space has a tendency to {\it simplify}
the flow, with the chaotic region expanding to fill the entire phase space. In what
follows we consider the details of the Martyna-Klein-Tuckerman oscillator :
$$
\{ \ \dot q = p \ ; \ \dot p = -q - \zeta p \ ; \ \dot \zeta = p^2 - 1 - \xi \zeta\ ; \
\dot \xi = \zeta^2 - 1 \ \} \ [ \ {\rm MKT} \ ]\cite{b6} \ .
$$
All of the numerical work described here was carried out with the classic fourth-order
Runge-Kutta integrator, mostly with a timestep of 0.001 . We consider six different aspects
of the MKT oscillator's phase-space flow, and show that all of them are fully consistent
with the ergodicity of that model.

\subsection{Moments of the Distribution Function}

\begin{figure}
\vspace{1 cm}
\includegraphics[width=4.0in,,angle=-90]{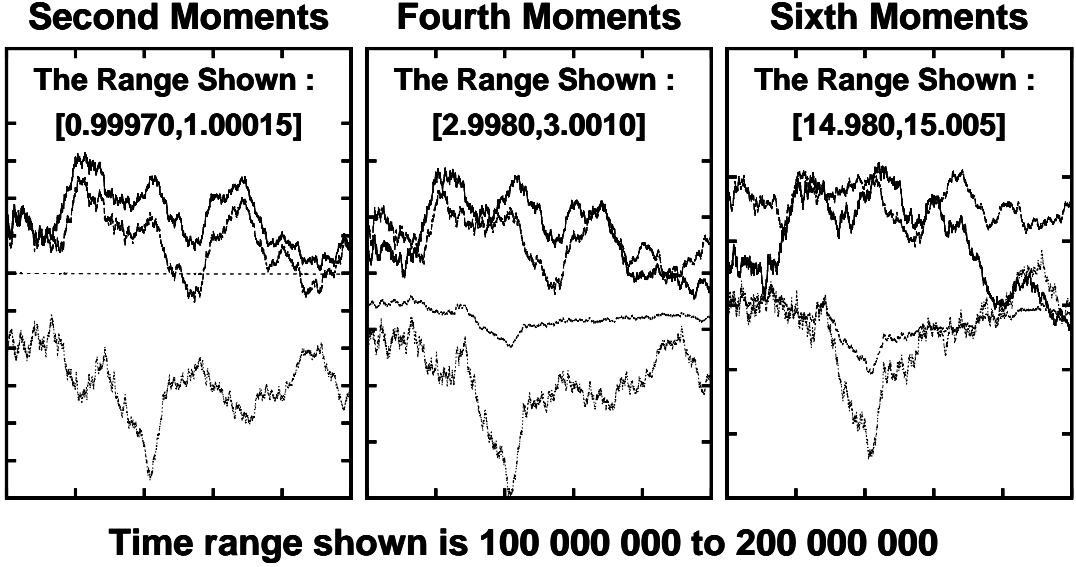}
\caption{
Typical variation of cumulative second, fourth, and sixth moments of the MKT oscillator.
The exact moments, 1, 3, and 15, are reproduced with four-figure accuracy for simulations
describing on the order of $10^8$ vibrations.
}
\end{figure}

If MKT dynamics is ergodic then its long-time-averaged distribution is {\it Gaussian} :
$$
f(q,p,\zeta,\xi) \propto e^{-q^2/2}e^{-p^2/2}e^{-\zeta^2/2}e^{-\xi^2/2} \ .
$$
The independence of the four variables implies that the second, fourth, and sixth
moments are equal to 1, 3, and 15 for each of them. {\bf Figure 1} compares the evolution
of all 12 of these  moments for the Martyna-Klein-Tuckerman model.  The moments are fully
consistent with the ergodic distribution.  The moments are reproduced to an accuracy of
about four significant figures during a simulation of $10^{12}$ timesteps. It should be
noted that because the last of the MKT equations, $\dot \xi = \zeta^2 - 1$ , forces the
longtime value of $\langle \ \zeta^2 \ \rangle $ to be unity, the ( numerical ) error of
that moment, of order $10^{-6}$ , is smaller than that for all the rest . For all
the other data, using a timestep of $dt = 0.001$ the single-step integration error is of
order $10^{-17}$ , about the same as double-precision roundoff error.  The number of
oscillations represented is about $10^8$ , quite consistent with a random error on the
order of the inverse square root of the number of independent sample oscillations.

\subsection{Chaoticity and the Largest Lyapunov Exponent}

Chaos is an essential ingredient of ergodicity. Chaos can be quantified by measuring the
evolution of Lyapunov instability, the ongoing tendency toward the exponentially-fast
separation of neighboring phase-space trajectories. A steady-state measurement of
Lyapunov instability can be implemented by forcing a tethered ``satellite'' trajectory
to follow the lead of a ``reference'' trajectory. Both reference and satellite follow
exactly the {\it same} motion equations but with the reference-to-satellite separation
continually constrained by rescaling its phase-space separation $\Delta$ at the end of
every timestep :
$$
\Delta_t \equiv r_s(t) - r_r(t) \ ; \ \Delta_{t+dt} \longrightarrow 
\Delta_{t+dt}\left[\frac{|\Delta_{t}|}{|\Delta_{t+dt}|}\right] \ .
$$
The rescaling of the separation, per unit time, defines the local Lyapunov exponent
$\lambda(t)$ :
$$
\lambda(t) \equiv (1/dt)\ln\left[\frac{|\Delta_{t+dt}|}{|\Delta_t|}\right]
\simeq (1/dt)\ln[ \ e^{\lambda dt} \ ] \equiv \lambda \ .
$$
The long-time-averaged value of this ``local'' time-dependent Lyapunov exponent is {\it the}
exponent $\lambda $ :
$$
\langle \ \lambda(t) \ \rangle \equiv \lambda \simeq
 \bigg \langle \ (1/dt)\ln \left[ \ \frac{|\Delta _{t+dt}|}{|\Delta _t|} \ \right] \ \bigg \rangle \ .
$$
The continuous limit $dt \rightarrow 0$ can be imposed by using an appropriate Lagrange
multiplier\cite{b13}.

\begin{figure}
\includegraphics[width=4.0in,,angle=-90]{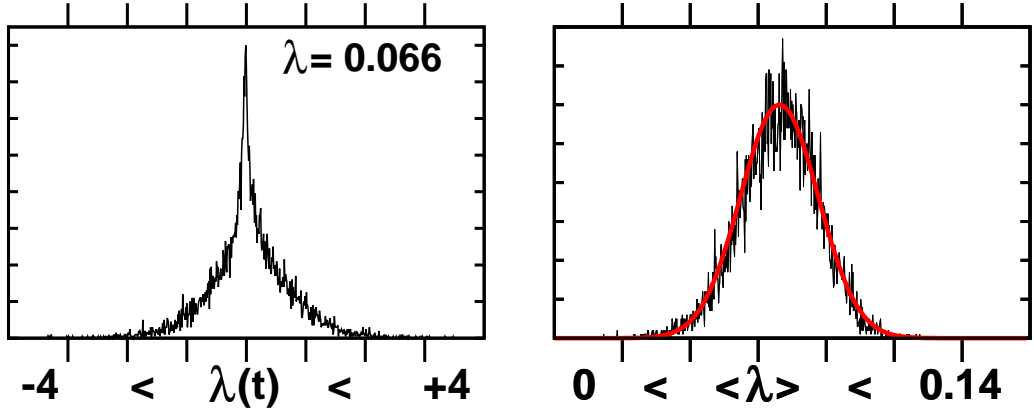}
\caption{
Instantaneous values of the Lyapunov exponent ( on the left ) and a histogram of integrated
million-timestep averages ( on the right ).  The Gaussian distribution at the right is drawn
with the observed mean and standard deviation and is a near-perfect fit to the data.  These
are probability densities so that the vertical scales are set by normalization. 
}
\end{figure}

{\bf Figure 2} compares a histogram of 10,000 values of the instantaneous ``local'' exponent 
$\lambda(t)$ , separated by 1000 timesteps with $dt = 0.001$ , to a histogram of integrated
averages.  Each averaged exponent represents one million timesteps, $\Delta t = 1000$ . The
reference-to-satellite offset vector $\Delta$ has a length 0.000001 . The time averages have
a mean and standard deviation : 
$$
\langle \ \lambda(t) \ \rangle_{\Delta t = 1000} = 0.066 \pm 0.011 \ .
$$
If the distribution were Gaussian the probability of finding a vanishing integrated exponent
in a million trials would be about $e^{-18} \simeq 10^{-8}$ .  The Gaussian shown in the
Figure leads to two conclusions: [ 1 ] one million timesteps are clearly sufficient
for the Central Limit Theorem to apply, so that [ 2 ] the likelihood of finding a
false-negative time average is indeed about one in 100 000 000.  For all three
of the time-reversible harmonic oscillator thermostat models, HH, JB, and MKT,
samples of one million time-averaged Lyapunov exponents were examined\cite{b11}.  The initial
conditions were chosen randomly from the four-dimensional stationary distributions.  The
data were consistent with chaos and with the absence of regular toroidal trajectories
( which would correspond to vanishing Lyapunov exponents ).  These Lyapunov exponent
investigations establish that the measure of any nonergodic component is less than 0.000001 .

\subsection{Analysis Near the Two MKT Fixed Points: (0,0,-1,+1) and (0,0,+1,-1)}

The MKT oscillator has two separated fixed points where the coordinate and momentum vanish,
$q=p=0$ .  The thermostat variables are $(\zeta,\xi) = (-1,+1) \ {\rm or} \ (+1,-1)$ .  The
apparent ergodicity of the oscillator implied by the Gaussian moments and the positive
Lyapunov exponent suggests that neither fixed point is stable.  To show that {\it both} are
actually exponentially unstable we linearize the equations of motion about the fixed points
by considering a perturbation vector $\delta = (\delta_q,\delta_p,\delta_\zeta,\delta_\xi)$ .

We begin with the fixed point singled out for analysis by Patra and Bhattacharya\cite{b9}
 $(q,p,\zeta,\xi)=(0,0,-1,+1)$ :
$$
\dot \delta_q = \delta_p \ ; \ \dot \delta_p =-\delta_q+\delta_p \ ; \
\dot \delta_\zeta = \delta_\xi -\delta_\zeta \ ; \ \dot \delta_\xi = - 2\delta_\zeta \ .
$$
Another time differentiation provides the separated equations of motion for the perturbations
in the $(q,p)$ and $(\zeta,\xi)$ planes.  The $(q,p)$ perturbations are linearly unstable, both
in the same manner :
$$
\ddot \delta_q = -\delta_q + \dot \delta_ q \ ; \
\ddot \delta_p = -\delta_p + \dot \delta_ p \ . 
$$
The perturbations in the $(\zeta, \xi)$ plane {\it are stable}, again in the same manner :
$$
\ddot \delta_\zeta = - 2\delta_\zeta - \dot \delta_\zeta \ ; \
\ddot \delta_\xi = -2\delta_\xi -\dot \delta_\xi \ .
$$
Evidently the $(\zeta,\xi)$ flow {\it toward} this fixed point complements
the corresponding Lyapunov-unstable exit flow in the $(q,p)$ plane.

Exactly similar analysis can be carried out at the other fixed point $(q,p,\zeta,\xi)=(0,0,+1,-1)$ :
$$
\dot \delta_q = \delta_p \ ; \ \dot \delta_p = -\delta_q - \delta_p \ ; \
\dot \delta_\zeta = -\delta_\xi +\delta_\zeta \ ; \ \dot \delta_\xi = + 2\delta_\zeta \ .
$$
This time perturbations $(\delta_q,\delta_p)$ in the $(q,p)$ plane are linearly {\it stable}
rather than unstable, and both in the same manner :
$$                         
\ddot \delta_q = -\delta_q - \dot \delta_q \ ; \ \ddot \delta_p = -\delta_p - \dot \delta_p \ .   
$$
In parallel, the perturbations in the $(\zeta, \xi)$ plane {\it are unstable} :
$$
\ddot \delta_\zeta = - 2\delta_\zeta + \dot \delta_\zeta \ ; \
\ddot \delta_\xi = -2\delta_\xi +\dot \delta_\xi \ .
$$
In summary, both the fixed points are exponentially unstable, with stable entrance and unstable
exit flows balancing in the steady state.

In addition to the mirror symmetry mentioned in the first Section all three sets of thermostated
oscillator equations exhibit a time-reversal symmetry in which the signs of $(p,\zeta,\xi)$ and
the time all change while that of the coordinate $q$ does not.  This symmetry implies that the
attractors and repellors change roles in the time-reversed dynamics, with damped stable oscillation
reversed to give unstable divergent oscillation and {\it vice versa}.  In either case the
oscillator equation $\ddot \delta_q = -\delta_q \pm\dot \delta_q$ corresponds to successive
amplitude changes larger or smaller by a factor 6.1 .  The parallel thermostat equation near the
$(\zeta,\xi)$ fixed points, $\ddot \delta_\zeta = -2\delta_\zeta \pm\dot \delta_\zeta$ corresponds
to successive amplitude changes of a factor of 3.3 at the control variable's turning points, smaller
because the characteristic frequency of these thermostat variables is greater.

\begin{figure}
\includegraphics[width=4.0in]{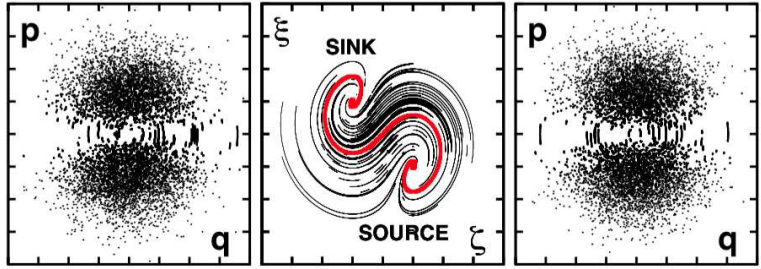}
\caption{
$(q,p)$ data are plotted for values close to the ``attractor'' ( left ) and ``repellor''
( right ) values of $(\zeta,\xi)$ =$(-1,+1)$ and $(+1,-1)$.  The middle panel shows $(\zeta,\xi)$
data for $(q,p)$ values close to $(0,0)$ . Both ``fixed points'' are exponentially unstable. The
ranges shown for all four variables are from -4 to +4.
}
\end{figure}

Evidently the $(q,p)$ flow {\it toward} the fixed point $(0,0)$ competes with the exit unstable
flow in the $(\zeta,\xi)$ plane.  One might expect that exponential divergence would overwhelm the
exponential slowing.  In order better to understand the fixed point flows we collect trajectory
points in three thin four-dimensional slabs centered on $(q,p)=(0,0)$ and on $(\zeta,\xi)=(-1,+1)$
and $(+1,-1)$ .  The thickness of the three slabs are all $10^{-5/2}$ .  See {\bf Figure 3} .  The
two cross-sectional views of $(q,p)$ look precisely similar at the two $(\zeta,\xi)$ ``fixed
points''.  While $q$ and $p$ are both small ( corresponding to a measure of 0.00001 ) the
$(\zeta,\xi)$ flow parallels the curve joining the two fixed points and emphasized in the center
of the Figure.  The relatively lengthy $(\zeta,\xi)$ excursions correspond to much smaller
$(q,p)$ tracks. In the thin slabs with $\zeta\xi \simeq -1$ , the coordinate changes by more than
a factor of six between crossings, and the amplitude of the $(q,p)$ motion is much less.  These
two effects are responsible for the misleading appearance of ``holes'' in the $(q,p)$ projections.
We will soon show that the density in the full four-dimensional $(q,p,\zeta,\xi)$ space is
actually completely uniform near both of the fixed points.  It is simply the jumps in $q$ coupled
with the slow  flow in $p$ that accounts for the low-density appearance emphasized by Patra and
Bhattacharya\cite{b9}.

\subsection{Exponential Motion Near the Fixed Points}

Near the two fixed points the flow is dominated by the source-to-sink S-curve shown in the central
panel of Figure 3 and well approximated by the $(\zeta,\xi)$ projection in the right panel of
{\bf Figure 4}.  It is educational to confirm the linear stability analysis by considering the
flow shown in Figure 4, starting very near the $(\zeta,\xi)$ ``source'' and $(q,p)$ ``sink'' :
$$
(q,p,\zeta,\xi) = (10^{-12},10^{-12},+1+10^{-12},-1-10^{-12}) .
$$
At the right in {\bf Figure 4} we plot the $(q,p)$ and $(\zeta,\xi)$ trajectories. From a visual
standpoint the $(\zeta,\xi)$ trajectory leaves the repellor at a time just past 40 and quickly
moves to the attactor, settling there near a time 60, and remaining there until just past 140.
During all that time the $(q,p)$ coordinates appear motionless. Their distance from the origin
corresponds to the heavy line in the left panel. The distance to the $(\zeta,\xi)$ repellor is
the light line there.  The medium left-panel line is the distance to the $(\zeta,\xi)$ attractor
reached near time 60.

After a time of 148 chaos ensues and the linear stability analysis visible in the left panel
no longer applies. The oscillator and thermostat plane trajectories in the right panel are typical and
show that the $(q,p)$ motion is slower and less vigorous than the $(\zeta,\xi)$ motion.

\begin{figure}
\includegraphics[width=4.0in,,angle=-90]{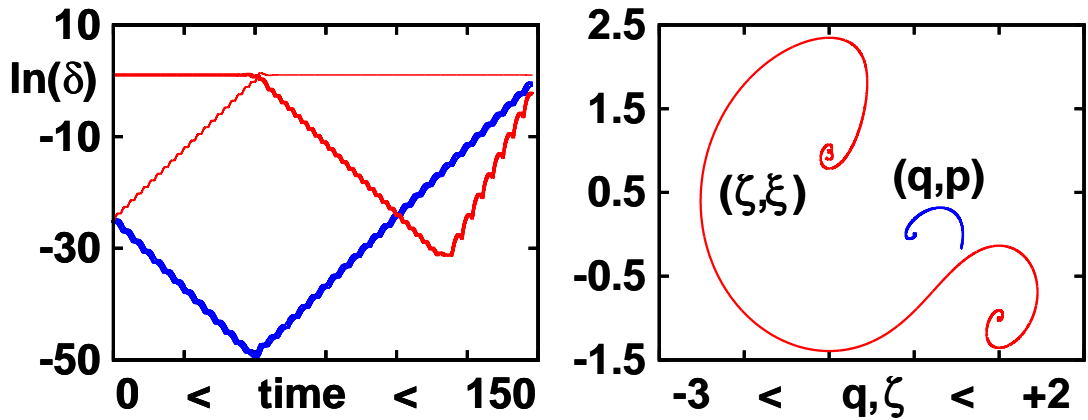}
\caption{
The left panel shows the $(q,p)$ and $(\zeta,\xi)$ separations from $(0,0)$ and $(\mp1,\pm1)$
for a simulation started very near (0,0,+1,-1) .  The heavy blue line shows the $(q,p)$ separation,
nearly invisible until a time of 140, then spiraling away from the origin while the light and
medium red lines show that the $(\zeta,\xi)$ track travels {\it from} the repellor ( along the
light line ) and {\it toward} the attractor ( medium ) by a time of 60, remaining near there
until visible chaos ensues at time 140 .
}
\end{figure}

\subsection{Probability Density Near the Two Fixed Points}

In {\bf Figure 5} we plot ( on logarithmic scales ) the number of points out of $10^{10}$ ( lower
curve ), $10^{11}$ , and $10^{12}$ ( upper curve ) lying within a distance $r$ of the two fixed
points, with $\ln (r)$ ranging from -5 to + 2.  Because $d\ln(r) = (dr/r)$ the density in
four-dimensional space should vary as $r^4$ rather than $r^3$ , which it does, very accurately.
The measure at the fixed points is equal to $(1/4\pi^2e)dqdp d\zeta d\xi$ .  Because the volume
of a four-dimensional sphere of radius $r$ is $(\pi^2r^4/2)$ the probability of finding a
trajectory point near one of the two fixed points within the smallest radius $r = e^{-5}$
is $e^{-21}/8 \simeq 9\times10^{-11}$ , explaining why such points occur rarely even on a
$10^{12}$-point trajectory, as is shown in the Figure.

\subsection{Grid-Based Measures}

\begin{figure}
\vspace{1 cm}
\includegraphics[width=2.5in,,angle=-90]{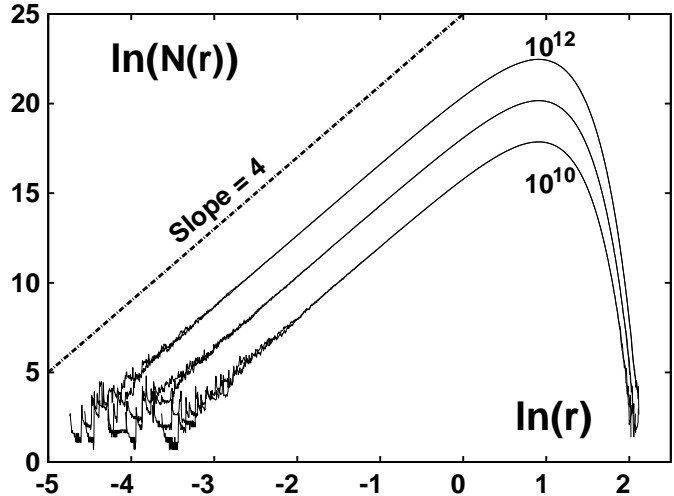}
\caption{
The number of trajectory points in the vicinity of the attractor and repellor are shown for
simulations of $10^{10}$ , $10^{11}$ , and $10^{12}$ points.  The two curves for each simulation
are only distinguishable for small $r$ where their fluctuations are visible.  The data are
consistent with a Gaussian density and show that the slope near 4 persists up to
spherically-averaged $r$ values of order unity.
}
\end{figure}

We can also confirm the Gaussian solution of the MKT equations by computing the measures
of 81 four-dimensional nonoverlapping balls of radius 0.50 arranged on a hypercubic 
$3\times 3\times 3\times 3$ grid centered on the origin.  The measures of the balls are
inversely proportional to the number of nonzero exponents.  The ball at the origin has none.
There are respectively 8, 24, 32, and 16 balls with their centers having 1, 2, 3, and 4
nonvanishing exponents.  A simulation with $10^{11}$ timesteps gave measures of 0.00719,
0.00445, 0.00276, $0.00170_6$, $0.00105_6$ for the five ball types. Each of these measures is close
to 0.620 times that of its predecessor, as expected for product measures of four independent
Gaussians. The two of the 81 balls centered on the dynamics' unstable fixed points show nothing
out of the ordinary in their measures relative to the other 22 two-exponent balls.  This
statistical test, which could be refined indefinitely in complexity, is, like all the rest,
consistent with ergodicity of the MKT oscillator equations.

\section{Summary}
The thermostated oscillator model introduced by Martyna, Klein, and Tuckerman, and explored
here in more detail, is one of three simple systems exhibiting a smooth Gaussian distribution
accompanied by a complex chaotic dynamics.  All three are important from the pedagogical
standpoint, and are also useful as thermostats generating all of Gibbs' canonical distribution.

Here we have summarized the ( compelling ) evidence for the ergodicity of the MKT oscillator
in order to close out the competition for the 2014 Snook Prize.  We have used six different
and independent methods to assess the ergodicity of the MKT oscillator: [1] the moments of the
Gaussian distribution; [2] the chaos, as opposed to regularity, of billions of independent
trajectories; [3] the instability of the flows near both fixed points; [4] the exponentially
growing separation from both fixed points; [5] the uniform probability density in the vicinity
of these unstable fixed points and [6] the expected relative measures within a set of 81
hyperspheres centered on the lattice nodes of a four-dimensional hypercubic lattice.

All of these  methods reach the same conclusion, that solutions of the coupled equations are
ergodic. We hope that this summary article will prove useful to investigators of ergodicity
in other simple dynamical systems.

In view of the very intricate Lyapunov instability of the Martyna-Klein-Tuckerman system this
ergodic Gaussian distribution is outstanding in its simplicity.  In view of the contributions
of Puneet Patra and Clint Sprott to the understanding of this problem we have divided the
2014 Snook Prize equally among ourselves and themselves.  We intend to formulate another
Snook Prize problem in the summer of 2015 and would be very grateful for suggestions from the
readers.  We thank Puneet Kumar Patra and Julien Clinton Sprott for helpful support and Ben
Leimkuhler and Mark Tuckerman for stimulating comments.

\pagebreak

\end{document}